\documentclass[pteplogo]{ptephy_v1}

\preprintnumber{}

\usepackage{amsmath,amssymb,bm}
\usepackage{graphicx}
\usepackage{booktabs,array,tabularx}
\usepackage{multirow}
\usepackage{adjustbox}
\usepackage{placeins}
\usepackage{xcolor}
\usepackage{comment}
\usepackage{url}
\usepackage{hyperref}

\graphicspath{{figures/}{./}}

\newcommand{\Heff}{H_{\mathrm{eff}}}
\newcommand{\eref}{E_{\mathrm{ref}}}
\newcommand{\psucc}{P_0}
\newcommand{\MeV}{\,\mathrm{MeV}}
\newcommand{\keV}{\,\mathrm{keV}}

\hypersetup{
  hidelinks,
  pdftitle={Ancilla-recycled RODEO filtering for nuclear scattering:
    A cross-platform benchmark on quantum processors},
  pdfauthor={
    Myeong-Hwan Mun, Jubin Park, Myung-Ki Cheoun, Eunja Ha
  }
}

\begin{document}

\title{
Ancilla-recycled RODEO filtering for nuclear scattering:
A cross-platform benchmark on quantum processors
}

\author[1]{Myeong-Hwan Mun}
\affil[1]{
Department of Physics,
Kyungpook National University,
Daegu 41566, Republic of Korea
}

\author[2]{Jubin Park\thanks{
Corresponding author.
E-mail: \texttt{honolov@ssu.ac.kr}
}}

\author[2]{Myung-Ki Cheoun\thanks{
Corresponding author.
E-mail: \texttt{cheoun@ssu.ac.kr}
}}

\affil[2]{
Department of Physics and
Origin of Matter and Evolution of Galaxies Institute,
Soongsil University,
Seoul 06978, Republic of Korea
}

\author[3]{Eunja Ha}
\affil[3]{
Department of Physics and
Research Institute for Natural Science,
Hanyang University,
Seoul 04763, Republic of Korea
}

\begin{abstract}
Quantum computation of trapped spectra for nuclear scattering requires
resource-efficient circuits and reliable energy inputs.
We present a cross-platform benchmark of ancilla-recycled RODEO
filtering for a four-dimensional effective Hamiltonian describing a trapped neutron--proton model.
Static circuits on IonQ Forte-1 and dynamic circuits on IBM Aachen
implement the same ideal ten-cycle all-zero accepted filter using twelve and 
three active qubits, respectively.
Mid-circuit measurement and reset reduce
the active qubit count by 75\% while retaining all controlled evolutions.
The interleaved IBM peak-center offset from the exact level is
$-0.017 \pm 0.507~\mathrm{keV}$, compared with
$-0.568 \pm 0.694~\mathrm{keV}$ for IonQ
(finite-shot bootstrap uncertainties).
At \(\omega=3.7~\mathrm{MeV}\), the modified effective range expansion
maps both centers to finite-confinement scattering inputs consistent
with the exact reference within propagated uncertainties.
Three IBM batches exhibit center variations beyond
finite-shot uncertainties (\(p=0.015\)), despite comparatively stable
postselection amplitudes.
Calibration-derived noisy simulations do not consistently reproduce
the measured centers.
Complete sampled support enables sample-based quantum diagonalization
to recover the exact effective-space level, whereas larger-space
simulator tests expose substantial errors from omitted configurations.
The calculation validates one finite-confinement input,
not a free-space phase shift.
This benchmark provides a reusable circuit and validation protocol
separating spectral precision, hardware repeatability, and
sampled-subspace completeness for multilevel
nuclear-scattering calculations.
\end{abstract}

\subjectindex{A61, A20, D00, D05}
\maketitle

\section{Introduction}
\label{sec:introduction}

Near-term quantum processors are constrained not only by coherence and
gate errors, but also by the number of simultaneously usable qubits
\cite{Preskill2018}. Auxiliary registers can therefore consume a
substantial fraction of the available hardware before the physical
system itself is enlarged. Mid-circuit measurement and reset provide a
way to exchange qubit width for sequential control: a measured ancilla
can be returned to \(\lvert0\rangle\) and reused later in the same
circuit \cite{DeCross2023}. This space--time tradeoff is particularly
relevant to spectral filters, phase-estimation variants, and other
ancilla-assisted algorithms whose deferred-measurement implementations
allocate separate ancillas to successive interrogation cycles.

Nuclear scattering is a natural application of such spectral methods.
Direct preparation of asymptotic scattering states remains difficult on
near-term hardware, while weak harmonic confinement replaces the
continuum by discrete positive-energy levels. Their trap-frequency
dependence is related to free-space phase shifts through the modified
effective range expansion (MERE)
\cite{Luu2010,Zhang2020,Wang2024NuclearScattering}. The quantum task
is then to determine trapped eigenenergies, whereas the interpolation in
energy and extrapolation to \(\omega\to0\) are performed classically.
This route complements variational, subspace, real-time, and
correlation-function approaches to scattering on quantum processors
\cite{Sharma2024PhaseShifts,Turro2024PhaseShifts,
Yusf2025ElasticScattering,Guo2026Scattering}.
In this formulation, the quantum output is a spectral input to a
subsequent continuum reconstruction. Resource reduction must therefore
be assessed together with the accuracy and repeatability of that input:
a circuit that uses fewer qubits is useful for scattering calculations
only if its energy estimates remain sufficiently controlled for the
classical analysis.

Wang \textit{et al.} demonstrated the weak-trap--RODEO--MERE framework
for model neutron--proton (\(np\)) and neutron--alpha (\(n\alpha\)) scattering on an ideal quantum
simulator \cite{Wang2024NuclearScattering}. The RODEO algorithm itself
filters an energy component through repeated ancilla-controlled time
evolutions \cite{Choi2021Rodeo}. Its canonical cycle admits immediate
ancilla measurement and reuse, and this form has been demonstrated on
IBM hardware for a one-qubit Hamiltonian \cite{Qian2024Rodeo}. By
contrast, the static Forte-1 realization used in the present work
deferred all measurements and unrolled the \(R=10\) cycles onto ten
ancillas. That static construction avoids mid-circuit operations but
uses twelve active qubits to interrogate a two-qubit Hamiltonian.
The ideal equivalence of static and recycled-ancilla filtering does not,
however, determine how their measured spectra compare in the presence
of different hardware error channels. In particular, it does not
establish whether a retained postselection signal implies an accurate
or repeatable energy estimate.

The central question of this work is whether the same nuclear
spectral primitive can be compressed to a three-qubit dynamic circuit
without sacrificing the trapped-energy information. We implement the
identical ideal \(R=10\) all-zero accepted filter in two
architecture-adapted forms: a static 12-qubit realization on IonQ
Forte-1 and an ancilla-recycled three-qubit realization on IBM Aachen.
The dynamic circuit reduces the active width by 75\%, while retaining
the same number of controlled evolutions and replacing nine ancillas by
ten mid-circuit measurements and nine resets. The Hamiltonian, initial
state, interrogation times, and query energies define a common spectral
benchmark. Our contribution is to connect this hardware implementation
to a validation protocol that separately examines circuit resources,
spectral-center accuracy, postselection attenuation, and the
reliability of the resulting nuclear-scattering input.

The compact four-dimensional Okubo--Lee--Suzuki (OLS) Hamiltonian
provides an exactly known nuclear reference for this comparison. It
retains selected trapped levels of a schematic neutron--proton
interaction, allowing the measured peak positions and sampled subspaces
to be assessed against the same effective-space spectrum. The reduced
space is classically tractable by construction; its role is to test the
hardware spectral step under controlled reference conditions before
repeating it over additional levels, confinement strengths, or larger
registers.

Three independent IBM batches, including an interleaved scan bracketed
by repeated central queries, are used to assess finite-shot precision
and batch-to-batch hardware repeatability. The interleaved dynamic scan
and the static IonQ scan give centers consistent with the exact level,
with sub-keV finite-shot uncertainties. Across the three IBM batches,
however, the centers are heterogeneous under their individual
finite-shot uncertainties (\(p=0.015\)), even though postselection
attenuation remains comparatively stable. We also compare
calibration-derived noisy simulations with the actual quantum
processing unit (QPU) data; these simulations do not consistently
reproduce the observed centers.
Together, the peak-fit, within-job, and between-batch checks distinguish
successful execution and statistical precision from spectral
repeatability. This distinction matters when energies obtained in
separate hardware executions are subsequently combined in a scattering
analysis.

We further audit the postselected model-space support by sample-based
quantum diagonalization (SQD)
\cite{Kanno2023QSCI,RobledoMoreno2024SQD}.
Both hardware implementations sample all four OLS configurations, so
the sampled Hamiltonian recovers the exact effective-space level.
This result diagnoses support completeness rather than the accuracy of
the noisy state or the fitted spectral center. Complementary
larger-space simulator tests show that near-complete observed support
can nevertheless leave substantial target-level errors. These tests
identify a separate validation requirement for extensions to larger
nuclear spaces: the energy sensitivity to missing configurations must
be examined rather than inferred from their sampling frequency alone.

To connect the hardware benchmark directly to nuclear scattering, we
map each measured \(E_2\) center at \(\omega=3.7~\mathrm{MeV}\)
to the finite-confinement scattering function
\(\mathcal K_{3.7}(E_c)\), expressed in \(\mathrm{fm}^{-1}\),
and propagate its finite-shot energy uncertainty to this MERE input. At the benchmark level, a
\(1~\mathrm{keV}\) center shift changes \(\mathcal K_{3.7}\) by
approximately \(2.89\times10^{-4}~\mathrm{fm}^{-1}\).
This conversion quantifies the consequence of spectral errors in the
quantity entering the continuum analysis. The present calculation
validates one finite-confinement input; a free-space determination
requires several levels and trap strengths. The hardware benchmark and
finite-confinement conversion therefore test the spectral component on
which a multilevel reconstruction depends.

The circuit and analysis protocol documented here provide a reusable
starting point for that reconstruction. The sequential
controlled-evolution, ancilla-measurement, and reset structure, together
with the recorded all-zero postselection rule, can be retained as the
Hamiltonian and target level are changed. The problem-dependent
evolutions must still be compiled and validated, but additional filter
cycles need not consume additional simultaneously active ancillas.
This three-qubit implementation already provides the starting point for
our subsequent study of factorized geometric-time RODEO, which extends
the calculation to four trapped levels at five confinement strengths
\cite{Mun2026FGRodeo}.
The present work establishes the cross-platform implementation and
batch-resolved spectral validation; the subsequent study addresses
further reduction of the evolution sequence and free-space phase-shift
reconstruction.

The remainder of this paper is organized as follows.
Section~\ref{sec:compression} describes the circuit construction and
the associated resource tradeoff. Section~\ref{sec:hardware} presents
the hardware spectra and repeatability tests, while
Section~\ref{sec:sqd} examines the sampled support and larger-space
diagnostics. Section~\ref{sec:nuclear_scattering_implications}
relates the measured level to the corresponding finite-confinement
scattering input. The conclusions are presented in
Section~\ref{sec:conclusions}, and the appendices provide numerical
and methodological details.

\section{Ancilla-recycled RODEO filtering}
\label{sec:compression}

\subsection{Conditional filtering and ancilla reuse}
\label{subsec:accepted_branch}

The four-dimensional trapped \(np\) OLS Hamiltonian \(\Heff\) is
encoded in two system qubits, \(n_s=\log_2 4=2\). Here \(n_s\)
counts the qubits representing the nuclear model space and excludes the
ancilla. The four computational-basis states label OLS model-space
vectors, rather than individual neutron or proton occupations.
Both implementations start from \(\lvert\psi_0\rangle=\lvert11\rangle\).
The Hamiltonian, eigenenergies, and computational-basis probabilities
are given in Appendix~\ref{app:benchmark}.
At query energy \(E\), the \(j\)-th RODEO cycle applies the controlled
evolution \cite{Choi2021Rodeo,Qian2024Rodeo}
\begin{equation}
U_j(E)=\exp[-i(\Heff-EI)t_j],
\label{eq:controlled_evolution}
\end{equation}
where \(I\) acts on the system register and \(t_j\) is the
interrogation time in units with \(\hbar=1\).
For an incoming system state \(\lvert\phi\rangle\), an ancilla
initialized in \(\lvert0\rangle_a\), and a controlled operation
activated by \(\lvert1\rangle_a\), the
Hadamard--controlled-\(U_j\)--Hadamard sequence gives
\begin{equation}
\lvert0\rangle_a\lvert\phi\rangle \longmapsto \frac{1}{2}\lvert0\rangle_a[I+U_j(E)]\lvert\phi\rangle + \frac{1}{2}\lvert1\rangle_a[I-U_j(E)]\lvert\phi\rangle.
\label{eq:cycle_joint_state}
\end{equation}

Postselecting the ancilla in \(\lvert0\rangle\) therefore produces
the conditional system operator
\begin{equation}
M_j(E)=\frac{I+U_j(E)}{2}.
\label{eq:rodeo_kraus}
\end{equation}
The unnormalized accepted state and the probability that all \(R\)
cycles succeed are
\begin{equation}
\begin{aligned}
\lvert\widetilde\psi_R(E)\rangle
&=M_R(E)\cdots M_1(E)\lvert\psi_0\rangle,\\
\psucc(E)&=\bigl\|\lvert\widetilde\psi_R(E)\rangle\bigr\|^2.
\end{aligned}
\label{eq:success_probability}
\end{equation}
For \(\psucc(E)>0\), the normalized postselected system state is
\(\lvert\widetilde\psi_R(E)\rangle/\sqrt{\psucc(E)}\).

In a static realization, each cycle has its own ancilla and all ancillas
are measured at the end. Each ancilla interacts with the system only
in its assigned cycle, so deferring its measurement leaves the
all-zero accepted branch in Eq.~\eqref{eq:success_probability}
unchanged. In the dynamic realization, the same ancilla is measured
after each cycle, its outcome is stored classically, and it is reset
before the next cycle. The system register is retained between cycles:
the filtered state remains in the system, while the classical record
stores the sequence of ancilla outcomes. Acceptance requires the entire
record to be zero, not merely the final measurement of the recycled
ancilla. After the last cycle, terminal system measurements provide the
postselected bit strings used in the SQD analysis of
Sec.~\ref{sec:sqd}.
The two realizations thus implement the same ideal all-zero accepted
filter for the same initial state and interrogation times. This
operator identity supplies the reference for the hardware comparison;
equality of the measured spectra is tested separately.

\subsection{Active-qubit resources and hardware costs}
\label{subsec:resource_tradeoff}

For the static unrolling and dynamic reuse constructions considered
here, the active-qubit requirements are
\(W_{\rm stat}=n_s+R\) and \(W_{\rm dyn}=n_s+1\), respectively.

For \(n_s=2\) and \(R=10\), the number of simultaneously active
qubits is reduced from 12 to 3, corresponding to a 75\% reduction in
circuit width. The resulting resource tradeoff is summarized in
Table~\ref{tab:resources}. The saved resource is the simultaneous
allocation of \(R-1\) ancillas: the \(R\) controlled evolutions and
\(R\) recorded ancilla outcomes remain necessary for this
implementation of the filter.

\begin{table}[!htbp]
\caption{
Algorithmic resources for static unrolling and dynamic ancilla reuse.
The number of controlled evolutions and recorded ancilla outcomes is
unchanged; the dynamic circuit trades \(R-1\) simultaneously allocated
ancillas for mid-circuit measurement and reset. MCM denotes mid-circuit
measurement. Reset counts refer to operations between RODEO cycles,
excluding initialization before the first cycle.
}
\label{tab:resources}
\centering
\small
\begin{adjustbox}{max width=\linewidth}
\begin{tabular}{lcc}
\toprule
Resource & Static & Dynamic \\
\midrule
Width & \(n_s+R\) & \(n_s+1\) \\
Active-qubit width (benchmark) & 12 & 3 \\
Ancillas & \(R\) & 1 \\
Controlled evolutions & \(R\) & \(R\) \\
Recorded ancilla outcomes & \(R\) & \(R\) \\
Ancilla measurement timing & terminal & mid-circuit \\
Resets & 0 & \(R-1\) \\
Main overhead & width & MCM/reset \\
\bottomrule
\end{tabular}
\end{adjustbox}
\end{table}

Ancilla reuse compresses circuit width, not the controlled-evolution
sequence. It introduces measurement and reset latency and additional
idle exposure of the system register. The submitted IBM circuits act
on physical qubits 153--155 and contain ten mid-circuit ancilla
measurements, nine resets, two terminal system measurements, and
112 controlled--Z (CZ) gates; their depths range from 447 to 453.

Query-dependent metrics for the compiled circuits are reported in Table~\ref{tab:app_metrics} of Appendix~C.
Corresponding native metrics on the
trapped-ion device are not directly comparable because the processor
architecture and compilation stack differ. We therefore compare the
common spectral observable rather than rank the platforms by gate
count or interpret the width reduction as an execution-time speedup.

The reusable element is the sequence of controlled evolution, ancilla
measurement, reset, and recorded-outcome postselection. Changing the
Hamiltonian or target level requires the corresponding evolutions and
initial state to be specified and validated, but does not change the
one-ancilla structure. At fixed system-register size, additional
interrogation cycles therefore increase the sequential work without
increasing the simultaneous ancilla requirement. This separates the
width cost of the filter from that of the nuclear register while leaving
controlled-evolution synthesis and hardware repeatability as explicit
requirements for extensions.

\subsection{Common interrogation times and scan protocols}
\label{subsec:scan_protocol}

Both implementations use \(R=10\) and the same deterministic,
symmetric Gaussian-quantile interrogation times,
\begin{equation}
t_j=\sigma\,\Phi^{-1}\!\left(\frac{j-1/2}{R}\right),
\label{eq:interrogation_times}
\end{equation}

where \(\sigma=21\,\mathrm{MeV}^{-1}\) and \(\Phi^{-1}\) is the
quantile function of the standard normal distribution.
The time values are listed in Table~\ref{tab:app_times} of Appendix~B.
Using the same realized time sequence fixes the ideal filter across
the two platforms and the independent batches, rather than introducing
a different time-sampling realization into each comparison.

The seven distinct query offsets are \(E-E_2=\{-40,-28,-16,0,+16,+28,+40\}\keV\).
IonQ and IBM Run~1 submit the queries in ascending order, whereas
IBM Run~2 submits them in descending order
\((+40,+28,+16,0,-16,-28,-40)\keV\).
IBM Run~3 submits the interleaved sequence
\((0,+16,-16,+28,-28,+40,-40,0)\keV\),
with central queries at the beginning and end of the submitted scan.
Each query circuit uses 500 shots. The two Run~3 central records are kept separate
for the within-job comparison and pooled only for the primary
seven-point fit, giving 1000 shots at the center. Thus the repeated
center is an additional control measurement, not a change in the
Hamiltonian or interrogation times.

The three IBM batches use the same archived instruction-set
architecture (ISA) circuits, rearranged only in submission order.
All primary results use raw counts without debiasing, dynamical
decoupling, twirling, or error mitigation.

The job identifiers and raw success counts are listed in Tables~\ref{tab:app_jobs} and \ref{tab:app_rawcounts} of Appendix~C.
This common circuit and scan definition
allows the center, attenuation, and repeatability checks to be applied
to the same spectral task.

\subsection{Spectral-center estimation and signal retention}
\label{subsec:spectral_estimation}

For
\(\lvert\psi_0\rangle=\sum_\lambda c_\lambda\lvert\lambda\rangle\),
with \(\Heff\lvert\lambda\rangle=E_\lambda\lvert\lambda\rangle\),
define the fixed-time response of one energy component as
\begin{equation}
f_{\bm t}(x)=\prod_{j=1}^{R}\cos^2(x\,t_j/2).
\end{equation}
Equation~\eqref{eq:success_probability} then gives the ideal spectral
response \(P_{\bm t}(E)\). The full response and the local two-parameter
peak model are
\begin{equation}
\begin{aligned}
P_{\bm t}(E)&=\sum_\lambda |c_\lambda|^2
                    f_{\bm t}(E-E_\lambda),\\
P_{\rm fit}(E;E_c,K)&=Kf_{\bm t}(E-E_c).
\end{aligned}
\label{eq:filter_response}
\end{equation}
The first expression includes all four effective-space eigenstates;
the second is the local model fitted over the target scan window.
Because the interrogation times are fixed, their explicit product
response is used in the primary fit. The Gaussian-averaged profile is
used as a line-shape cross-check, as detailed in
Appendix~\ref{app:fitting} and Table~\ref{tab:app_fitmodels}.

For \(N_{\rm sh}\) shots at a given query, the measured probability is
\begin{equation}
\widehat P_0(E)=\frac{N_0(E)}{N_{\rm sh}},
\label{eq:measured_success_probability}
\end{equation}
where \(N_0(E)\) counts records in which all \(R\) ancilla outcomes
are zero. Fits use binomial standard errors, with \(E_c\) and \(K\)
as free parameters. The fitted center \(E_c\) estimates the trapped
eigenenergy, whereas \(K/K_{\rm loc}^{\rm ex}\) measures the retained
amplitude relative to the noiseless fixed-time response. The common
reference \(K_{\rm loc}^{\rm ex}\simeq0.894\) is obtained by fitting
the full noiseless response at the same seven query energies. It is
therefore not defined as the isolated target-state overlap, even though
it is numerically close to
\(\lvert\langle E_2\vert\psi_0\rangle\rvert^2=0.8941925\).

Within this local description, attenuation primarily changes the number
of accepted shots, while a center displacement changes the energy
entering the MERE. Reporting both fit parameters separates signal
retention from spectral-center accuracy rather than using the
postselection yield as a proxy for both. Finite-shot center
uncertainties are obtained from 
%3000 successful parametric-bootstrap refits: %
2000 successful parametric-bootstrap refits for IonQ and 3000 for IBM: 
counts are drawn 
%from the fitted binomial probabilities 
from binomial distributions with the observed success fractions 
using the original shot allocations, and the same model is refitted.
These uncertainties describe within-batch finite-shot precision; the
independent-batch comparisons in Sec.~\ref{sec:hardware} assess
repeatability separately.

\begin{figure}
\centering
\includegraphics[width=0.92\textwidth]{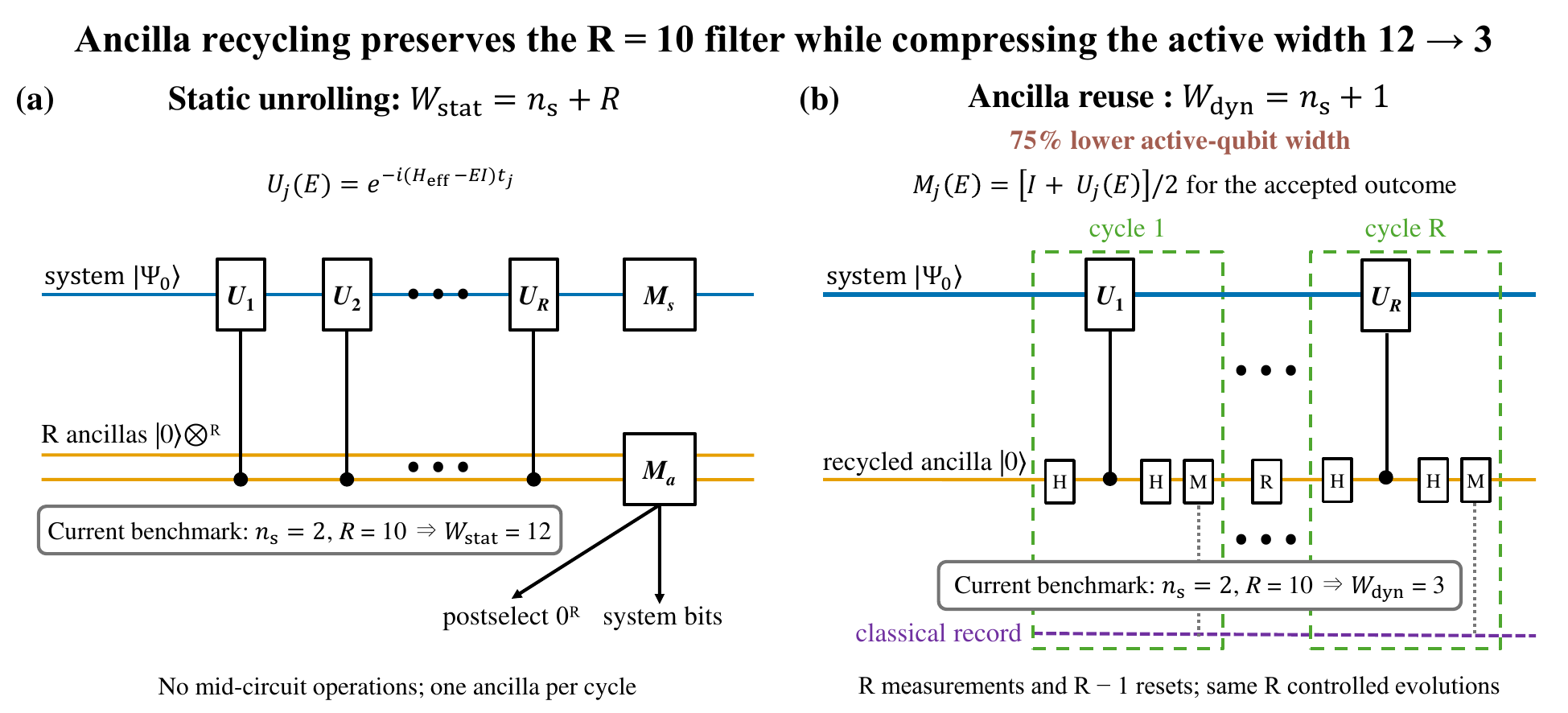}
\par\vspace{1mm}
\begin{minipage}[t]{0.49\textwidth}
\centering
\includegraphics[width=\linewidth]{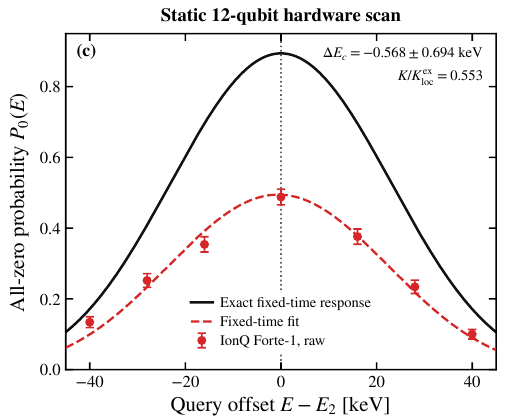}
\end{minipage}
\hfill
\begin{minipage}[t]{0.49\textwidth}
\centering
\includegraphics[width=\linewidth]{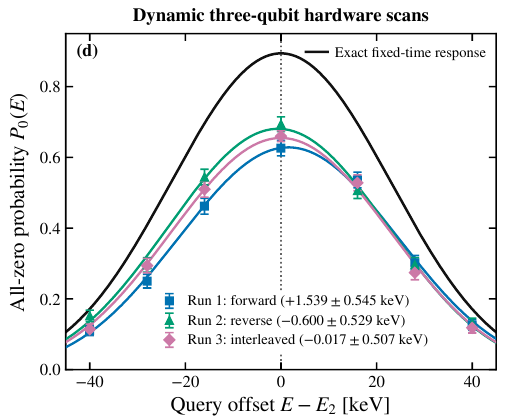}
\end{minipage}
\caption{
Qubit-width compression of the ideal all-zero accepted branch of the
\(R=10\) RODEO filter and the resulting hardware spectra.
(a) Static realization with one ancilla per cycle, requiring
\(W_{\rm stat}=n_s+R\) active qubits.
The initial and final Hadamard gates on the static ancillas are omitted
for visual clarity. Here \(M_s\) and \(M_a\) denote terminal system-
and ancilla-register measurements, respectively.
(b) Dynamic realization with one measured, reset, and recycled ancilla,
requiring \(W_{\rm dyn}=n_s+1\) active qubits. For \(n_s=2\) and
\(R=10\), the width is reduced from 12 to 3 while the ten controlled
evolutions are retained. Here \(M\) and \(R\) denote mid-circuit
measurement and reset, respectively. The terminal measurement of the
two-qubit system register, which supplies the bit strings used for the
SQD analysis, is omitted for visual clarity.
(c) Raw IonQ Forte-1 scan and fixed-time fit.
(d) Three raw IBM Aachen scans using forward, reverse, and interleaved
query protocols. Curves are the corresponding fixed-time fits and the
black curve is the noiseless response. Error bars are binomial standard
errors; the dotted lines mark the exact \(E_2=17.134039\MeV\).
}
\label{fig:compression}
\end{figure}

Figure~\ref{fig:compression} displays the static and dynamic circuit
structures together with their measured spectra. The common ideal
filter, explicit resource accounting, and shared spectral estimator
provide the basis for testing whether the reduced-width implementation
retains the nuclear-energy input on hardware.

\section{Cross-platform hardware benchmark}
\label{sec:hardware}

\subsection{Trapped-level reference and hardware spectra}
\label{subsec:hardware_spectra}

We use the model \(0^+\) square-well \(np\) system of
Ref.~\cite{Wang2024NuclearScattering} at \(\omega=3.7\MeV\).
The four-dimensional OLS effective Hamiltonian
\cite{Okubo1954,SuzukiLee1980} retains the trapped levels
\(E_1\)--\(E_4\); the target is
$\eref=E_2=17.134039\MeV$.
This positive-energy level supplies one finite-\(\omega\) input to
MERE, whose conversion is examined in
Sec.~\ref{sec:nuclear_scattering_implications}. Here the known
reference level and the common scan definition of
Sec.~\ref{sec:compression} allow the measured spectra to be compared
without changing the underlying nuclear problem. The benchmark tests
local spectral estimation around a known level, rather than a blind
search for an unknown eigenvalue.

Figure~\ref{fig:compression}(c,d) shows the raw static and dynamic
spectra. The hardware results obtained with the common fixed-time
estimator are summarized in Table~\ref{tab:hardware}. All center
uncertainties quoted in this section are finite-shot
parametric-bootstrap standard deviations unless otherwise specified;
they do not include the observed variation between independent batches.

\begin{table}[!htbp]
\caption{
Raw-count hardware results for the trapped \(E_2\) benchmark. Each
query circuit uses 500 shots; the interleaved IBM central point pools
two 500 shot records. Here \(\Delta E_c=E_c-E_2\),
\(\sigma_E\) is the finite-shot parametric-bootstrap uncertainty,
and \(K/K_{\rm loc}^{\rm ex}\) uses the common noiseless fixed-time
reference. The final column gives the observed SQD support relative to
the complete OLS space. The quoted uncertainties describe within-batch
finite-shot precision, not the full processor error budget.
}
\label{tab:hardware}
\centering
\small
\setlength{\tabcolsep}{4.4pt}
\begin{adjustbox}{max width=\linewidth}
\begin{tabular}{llcccccc}
\toprule
Data set & Protocol & Active qubits &
\(\widehat P_0(E_2)\) & \(\Delta E_c\) (keV) & \(\sigma_E\) (keV)&
\(K/K_{\rm loc}^{\rm ex}\) & \(|S|/4\) \\
\midrule
IonQ Forte-1 & \shortstack[l]{static,\\ascending} & 12 & 0.488 &
\(-0.568\) & 0.694 & 0.553 & \(4/4\) \\
IBM Aachen 1 & \shortstack[l]{dynamic,\\forward} & 3 & 0.626 &
\(+1.539\) & 0.545 & 0.703 & \(4/4\) \\
IBM Aachen 2 & \shortstack[l]{dynamic,\\reverse} & 3 & 0.694 &
\(-0.600\) & 0.529 & 0.761 & \(4/4\) \\
IBM Aachen 3 & \shortstack[l]{dynamic,\\interleaved} & 3 & 0.660 &
\(-0.017\) & 0.507 & 0.732 & \(4/4\) \\
\bottomrule
\end{tabular}
\end{adjustbox}
\end{table}

The static Forte-1 scan gives 
$\Delta E_c^{\rm IonQ}=-0.568\pm0.694_{\rm stat}\keV$,
so no center displacement is statistically resolved. The dynamic
interleaved scan gives
$\Delta E_c^{\rm IBM,I}=-0.017\pm0.507_{\rm stat}\keV$,
which is also consistent with the exact level. Thus the three-qubit
realization resolves the target center with sub-keV finite-shot
precision comparable to the 12-qubit static result. Both implementations
also retain complete four-configuration OLS support, whose separate
significance is examined in Sec.~\ref{sec:sqd}.

These results demonstrate that the ten-cycle nuclear spectral filter
can be executed with one recycled ancilla while retaining an accurate
trapped-energy estimate in the controlled interleaved batch. The
numerically smaller uncertainty in that batch is not, by itself, an
intrinsic precision gain from recycling: the processors and scan
protocols differ, and the additional central query gives Run~3 a
larger total shot allocation. The common comparison is the preservation
of the spectral input under the reduced active-qubit requirement.

\subsection{Batch-resolved repeatability and postselection retention}
\label{subsec:batch_repeatability}

The three IBM batches yield
$\Delta E_c^{\rm F}=+1.539\pm0.545\keV$,
$\Delta E_c^{\rm R}=-0.600\pm0.529\keV$,
$\Delta E_c^{\rm I}=-0.017\pm0.507\keV$.
They use the same archived query circuits but different submission
orders, as described in Sec.~\ref{subsec:scan_protocol}. To compare
the centers against a common-center hypothesis, let
\(x_i=\Delta E_{c,i}\) and \(s_i=\sigma_{E,i}\). The
inverse-variance mean and Cochran statistic are
\begin{equation}
\bar x=\frac{\sum_{i=1}^{3}x_i/s_i^2}{\sum_{i=1}^{3}1/s_i^2},
\qquad
Q=\sum_{i=1}^{3}\frac{(x_i-\bar x)^2}{s_i^2}.
\label{eq:app_cochran}
\end{equation}
Using the individual finite-shot variances gives \(Q=8.45\) for
two degrees of freedom, with \(p=0.0146\), or \(0.015\) when
rounded. The center variation is therefore larger than described by
the within-batch finite-shot uncertainties under this comparison.
A pooled shot-only uncertainty would not constitute a complete
processor error budget.

The forward-run shift is not shared by all three batches, so these data
do not establish a common, reproducible processor bias. Since each
submission order is represented by one independent batch, order
dependence and batch dependence cannot be separated by this dataset.
The comparison diagnoses an additional repeatability requirement; it
does not identify a unique source of the variation. With only three
IBM batches, \(Q\) is used as a consistency diagnostic rather than
to construct a complete processor-error model. Repeatability is
quantified only on IBM because only one independent Forte-1 batch was
available.

\begin{figure}[!htbp]
\centering
\includegraphics[width=0.49\textwidth]{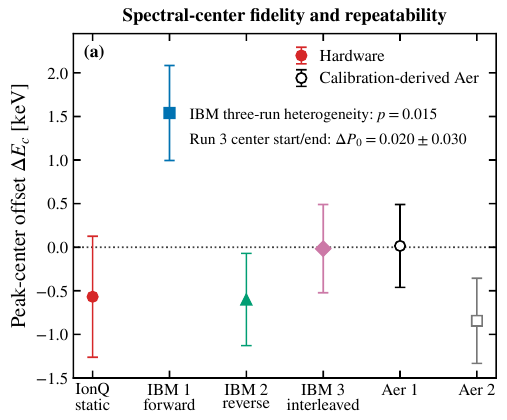}
\hfill
\includegraphics[width=0.49\textwidth]{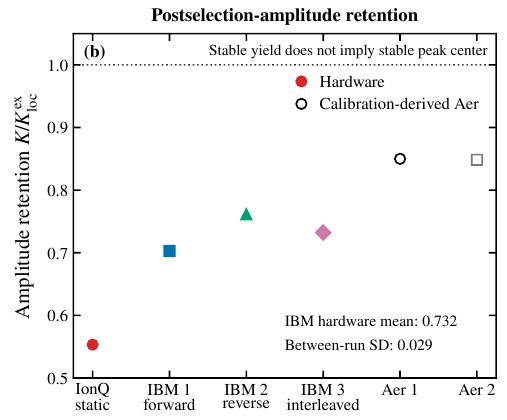}
\caption{
Separate validation observables for the trapped-energy filter.
(a) Peak-center offsets from the static IonQ batch, three dynamic IBM
batches, and calibration-derived IBM Aer simulations. Filled symbols
are QPU data and open symbols are noisy simulations; error bars are
finite-shot bootstrap uncertainties.
(b) Exact-reference-normalized amplitude retention for the same data.
The IBM amplitudes are comparatively stable even though the fitted
centers are heterogeneous under their finite-shot uncertainties.
The displayed between-run amplitude standard deviation (SD) is descriptive;
it is not an uncertainty on the spectral center.
}
\label{fig:validation}
\end{figure}

Postselection attenuation is comparatively stable across the same IBM
records. The amplitude ratios \(K/K_{\rm loc}^{\rm ex}\) are
0.703, 0.761, and 0.732, with mean 0.732 and between-run standard
deviation (SD) 0.029, whereas the fitted centers span more than
\(2\keV\). These amplitude summaries are descriptive, but their
contrast with the center test is operationally important: similar
accepted-event yields do not establish a common spectral center.
Figure~\ref{fig:validation} displays both quantities separately,
including the noisy-simulator results discussed below.

\subsection{Line-shape sensitivity and within-job controls}
\label{subsec:hardware_controls}

The repeatability comparison is supplemented by checks of the
line-shape model and the treatment of the interleaved central records.
Table~\ref{tab:app_fitmodels} compares the fixed-time model with the
Gaussian-averaged profile defined in Appendix~\ref{app:fitting}.
Figure~\ref{fig:app_fitcontrols}(a)
shows the corresponding fitted centers.

\begin{table}[!htbp]
\caption{
Fixed-time \((t)\) and Gaussian-averaged \((G)\) center fits to
the same hardware records. The uncertainties are parametric-bootstrap
standard deviations. The fixed-time reduced chi-square is
\(\chi_\nu^{2(t)}\), with \(\nu=7-2=5\) for the seven-point,
two-parameter fits. The final column uses the fixed-time amplitude.
The two models use the same counts and do not provide independent
energy measurements.
}
\label{tab:app_fitmodels}
\centering
\small
\begin{adjustbox}{max width=\linewidth}

\begin{tabular}{lcccccc}
\toprule
Data set & \(\Delta E_c^{(t)}\) (keV) & \(\sigma_E^{(t)}\)  (keV)&
\(\chi_\nu^{2(t)}\) & \(\Delta E_c^{(G)}\) (keV)&
\(\sigma_E^{(G)}\) (keV) & \(K/K_{\rm loc}^{\rm ex}\) \\

\midrule
IonQ Forte-1 & \(-0.568\) & 0.694 & 2.103 & \(-0.523\) & 0.724 & 0.553 \\
IBM Run 1 & \(+1.539\) & 0.545 & 0.509 & \(+1.640\) & 0.578 & 0.703 \\
IBM Run 2 & \(-0.600\) & 0.529 & 0.704 & \(-0.641\) & 0.534 & 0.761 \\
IBM Run 3 & \(-0.017\) & 0.507 & 0.665 & \(+0.005\) & 0.541 & 0.732 \\
\bottomrule
\end{tabular}
\end{adjustbox}
\end{table}

\begin{figure}[!htbp]
\centering
\includegraphics[width=0.93\textwidth]{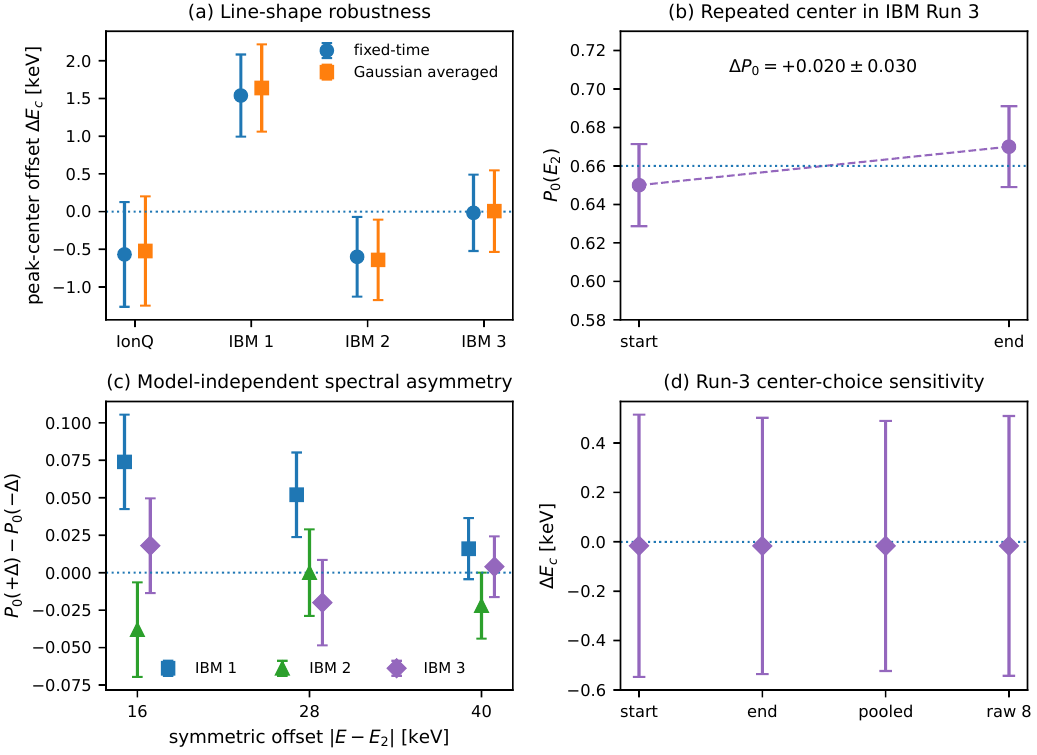}
\caption{
Peak-fit and internal-control diagnostics.
(a) Fixed-time and Gaussian-averaged center fits to the same counts.
(b) Repeated central-query probabilities in IBM Run~3.
(c) Raw symmetric-pair differences \(A_i(\Delta)\) defined in
Eq.~\eqref{eq:hardware_asymmetry}.
(d) Run~3 center obtained using the first central record, the final
central record, the pooled central record, or all eight query records
included without pooling. Panels (a) and (d) compare analyses of shared
data rather than independent measurements. The internal checks
complement, but do not replace, the separate-batch test in
Eq.~\eqref{eq:app_cochran}.
}
\label{fig:app_fitcontrols}
\end{figure}

The two profiles give the same qualitative pattern: the forward IBM
scan is shifted positively, while the static and interleaved centers
are consistent with the exact reference. At the reported precision,
changing the profile moves a center by at most \(0.101\keV\),
less than the corresponding finite-shot uncertainty. Thus this
particular line-shape choice does not account for the observed
between-batch spread. The IonQ dataset has the largest reduced
chi-square of the four fixed-time fits. Agreement of the two center
estimates is therefore a specific line-shape sensitivity check, not a
complete model-error estimate. All primary results continue to use the
explicit fixed-time response.

For Run~3, the first and last submitted central queries yield
325/500 and 335/500 successes, respectively. Their difference is
$\widehat P_0^{\rm end}(E_2)-\widehat P_0^{\rm start}(E_2)
=0.020\pm0.030$,
so this repeated-center test resolves no change between the two records
[Fig.~\ref{fig:app_fitcontrols}(b)]. A binomial fit to all eight raw
query records with an additional linear submission-position term gives
a likelihood-ratio-test \(p=0.969\). No linear
submission-position dependence is detected by that test. These checks
probe the recorded within-job consistency; they do not establish that
all time-dependent hardware errors are absent or that interleaving
causally removes the shifts seen in the other batches.

The symmetric query grid also supplies a check independent of the
line-shape fit. For a positive query offset \(\Delta\), define
\begin{equation}
A_i(\Delta)=\widehat P_{0,i}(E_2+\Delta)
            -\widehat P_{0,i}(E_2-\Delta),
\qquad \Delta\in\{16,28,40\}\keV,
\label{eq:hardware_asymmetry}
\end{equation}
where \(i\) labels the IBM batch.
Figure~\ref{fig:app_fitcontrols}(c) shows these raw probability
differences. The forward scan has positive central estimates at the two
inner pairs, consistent in sign with its fitted positive displacement;
the other batches do not reproduce the same pattern. These
pointwise comparisons provide a direct view of spectral asymmetry, not
an independent global significance test or a determination of its
hardware origin.

Finally, Fig.~\ref{fig:app_fitcontrols}(d) compares the Run~3 center
obtained using the first central record, the last record, their pooled
count, or all eight records without pooling. The displayed centers
remain close compared with their finite-shot uncertainties. This
checks that the near-zero interleaved center is not an artifact of
choosing one of its two central records. Since the fits share the same
off-center counts, they are sensitivity checks on one dataset, not
additional hardware repetitions.

\subsection{Comparison with calibration-derived noisy simulations}
\label{subsec:hardware_aer}
Calibration-derived Aer simulations associated with the first two IBM
campaigns give
$\Delta E_c^{\rm Aer,1}=+0.015\pm0.475\keV$ and
$\Delta E_c^{\rm Aer,2}=-0.845\pm0.488\keV$.
The second simulated center is compatible with the reverse hardware
batch within the quoted uncertainties, whereas the first does not
reproduce the positive center of the forward batch. The simulated
amplitude ratios are 0.850 and 0.848, compared with 0.703 and 0.761
for the corresponding hardware records. Thus the simulated amplitudes
are higher in both comparisons even though agreement of the centers
differs (see Fig.~\ref{fig:validation}). The underlying simulator
results are recorded in Appendix~\ref{app:diagnostics}.

These comparisons show why a calibration-derived noisy simulation and
independent hardware execution serve different validation roles. For
these datasets, the simulated response does not consistently describe
both the center and the attenuation. In particular, a small simulated
center error cannot substitute for a measurement of hardware
repeatability. The comparisons identify the limitation of the available
noise-model predictions without assigning the discrepancies to a
specific unmeasured noise mechanism.

\subsection{Implications for repeated spectral estimation}
\label{subsec:hardware_implications}

The hardware benchmark establishes two complementary results. First,
the ten-cycle filter can retain a trapped-energy input consistent with
the exact reference in the interleaved implementation while reducing
the active width from twelve to three qubits. Second, neither a retained
postselection signal nor a sub-keV finite-shot uncertainty is sufficient
to characterize the reliability of repeated spectral estimation. The
separate center, attenuation, line-shape, and repeatability checks make
this distinction explicit using the same nuclear reference problem.

The dynamic result is therefore not a processor ranking or evidence of
a reduced number of controlled evolutions. Its transferable contribution
is a reduced-width circuit implementation accompanied by measured
criteria for evaluating its spectral output. For subsequent multilevel
or multi-frequency calculations, this supplies a tested starting point
for the spectral step and identifies the checks needed before combining
energies from separate hardware executions. The sampled-support and
finite-confinement scattering analyses in the following sections examine
two further aspects of that output without replacing the center and
repeatability tests reported here.

\FloatBarrier
\section{Sampled-support validation with SQD}
\label{sec:sqd}

The purpose of SQD in the present calculation is not to refine the
RODEO peak-center estimate or to reconstruct the noisy quantum state.
Rather, it provides a support-completeness audit of the postselected
hardware records. Because RODEO retains only shots with an all-zero
ancilla record, hardware noise, mid-circuit measurement, and reset may
suppress system configurations that are important for representing the
target eigenstate. We therefore use SQD to test whether the sampled
system configurations span a sufficiently complete subspace to recover
the target Ritz eigenvalue. In the present four-dimensional benchmark,
complete support makes this recovery an identity, whereas the
incomplete-support and larger-space tests quantify the potential
projection bias relevant to future nuclear calculations.

\subsection{Postselected support and the hardware result}
\label{subsec:sqd_hardware_support}

Here the observed support \(S\) is the set of distinct system bit
strings recorded at least once among the accepted shots at the central
query. These are the terminal measurements already collected with the
RODEO scan; no additional quantum circuit is required for this audit.
Let \(V_S\) contain the corresponding computational-basis vectors.
SQD constructs
\begin{equation}
H_S=V_S^\dagger\Heff V_S
\label{eq:sqd}
\end{equation}
and diagonalizes this matrix classically
\cite{Kanno2023QSCI,RobledoMoreno2024SQD}.
The measurements select the basis, while the matrix elements come from
the known Hamiltonian; measured configuration frequencies are not used
as matrix elements. A Ritz eigenvalue is an eigenvalue of this
restricted Hamiltonian.

Table~\ref{tab:app_systemcounts} shows that every central-query dataset
contains \(S_{\rm full}=\{00,01,10,11\}\), including the two separate
Run~3 central datasets. Hence \(H_S\) is the full \(\Heff\), up to a
basis permutation, and its target eigenvalue is exactly \(E_2\).
No model-space configuration is missing from the observed support at
the reported shot counts.

\begin{table}[!htbp]
\caption{
Postselected system counts at the central query. The four configuration
counts in each row sum to the number \(N_0\) of all-zero ancilla
records. The Run~3 pooled row combines the first and final central
records; it is not an additional hardware execution. The bit strings
label OLS basis vectors, not neutron or proton occupations.
}
\label{tab:app_systemcounts}
\centering
\small
\begin{adjustbox}{max width=\linewidth}
\begin{tabular}{lrrrrr}
\toprule
Data set & \(N_0\) & \(11\) & \(10\) & \(00\) & \(01\) \\
\midrule
IonQ Forte-1 & 244 & 178 & 49 & 7 & 10 \\
IBM Run 1 & 313 & 256 & 29 & 16 & 12 \\
IBM Run 2 & 347 & 297 & 16 & 25 & 9 \\
IBM Run 3 start & 325 & 276 & 24 & 21 & 4 \\
IBM Run 3 end & 335 & 290 & 20 & 20 & 5 \\
IBM Run 3 pooled & 660 & 566 & 44 & 41 & 9 \\
\bottomrule
\end{tabular}
\end{adjustbox}
\end{table}

This result does not independently certify the prepared state or the
RODEO peak center. Finite sampling can leave configurations unobserved,
and measurement errors can also populate configurations. Indeed, the
forward IBM batch has complete support even though its fitted center
is displaced. SQD therefore identifies the absence of sampled-basis
projection error in this four-dimensional space; it does not correct
\(E_c\). The scattering conversion in
Sec.~\ref{sec:nuclear_scattering_implications} uses the measured
peak centers, not the exact SQD eigenvalue.

\subsection{What incomplete support would change}
\label{subsec:sqd_support_sensitivity}

Restricting the same Hamiltonian to \(\{11,10\}\) or
\(\{11,00\}\) gives target-related Ritz errors of
\(-50.1\keV\) and \(+572.8\keV\), respectively. These are
classical projection errors induced by omitting configurations, not
additional uncertainties of the RODEO fit. The Run~3 end dataset has
equal counts for \(10\) and \(00\), illustrating why sampling
frequency alone cannot determine their importance to the target energy.
The frequency-ranked trajectories and tie convention are documented in
Appendix~\ref{app:diagnostics}.

The larger-space simulator tests in the same appendix separate
oscillator-basis truncation from projection onto sampled configurations.
For example, observing 63 of 64 configurations still leaves a
\(+1.5374\MeV\) target-related projection error in the displayed
realization. These are sensitivity tests, not additional QPU results
or an estimate of how frequently such errors occur. Their implication
for extending the recycled-ancilla circuit is specific: increasing the
nuclear register requires checking which configurations were sampled
and how omissions affect the energy, in addition to checking the
spectral center and hardware repeatability. The present hardware data
pass the observed-support completeness check; the separate spectral
checks of Sec.~\ref{sec:hardware} remain necessary.

\FloatBarrier

\section{Nuclear-scattering implications of the measured trapped level}
\label{sec:nuclear_scattering_implications}

\subsection{From a trapped level to a scattering input}
\label{subsec:scattering_input}

The measured \(E_2\) is a positive-energy level of the relative
\(np\) Hamiltonian in a harmonic trap, rather than an eigenvalue of
an arbitrary test matrix. The underlying square well has a shallow
bound state \(B=2.22002\MeV\) and free-space scattering length
\(a_s=5.20134~\mathrm{fm}\); these characterize the input
interaction and are not extracted from the single QPU level. The
square-well matching and analytic reference are given in
Appendix~\ref{app:scattering}. The purpose of the following conversion
is to determine how the measured spectral accuracy translates into the
quantity supplied to a trapped-spectrum scattering analysis.

For an \(S\) wave, the trap quantization expression entering the
MERE is evaluated at the trapped eigenenergy
\cite{Luu2010,Zhang2020,Wang2024NuclearScattering}. We write it here
with the unit conversion explicit:
\begin{equation}
\mathcal K_\omega(E)
=-\frac{\sqrt{4\mu\omega}}{\hbar c}\,
\frac{\Gamma\!\left(\frac34-\frac{E}{2\omega}\right)}
     {\Gamma\!\left(\frac14-\frac{E}{2\omega}\right)}.
\label{eq:MERE_single_level}
\end{equation}
As in the preceding sections, the trap parameter \(\omega\) is
expressed as an energy in MeV. With \(E\), \(\omega\), and the
reduced mass \(\mu=469.460~\mathrm{MeV}\) expressed in energy
units, \(\hbar c=197.326980~\mathrm{MeV\,fm}\) gives
\(\mathcal K_\omega\) in \(\mathrm{fm}^{-1}\). The associated
wave number is \(k(E)=\sqrt{2\mu E}/(\hbar c)\). This convention
is the natural-unit quantity \(p\cot\delta_{0,\omega}\) divided
by \(\hbar c\), with \(p=\sqrt{2\mu E}\).

Evaluated at a trapped level, Eq.~\eqref{eq:MERE_single_level}
provides one finite-confinement scattering input. At
\(\omega=3.7\MeV\), the exact benchmark level gives
\(\mathcal K_{3.7}^{\rm exact}=0.13436~\mathrm{fm}^{-1}\).
For the hardware conversion we use the fitted centers \(E_c\) of
Sec.~\ref{sec:hardware}, not the exact eigenvalue recovered by
complete-support SQD. This keeps the propagated hardware error distinct
from the support-completeness identity in Sec.~\ref{sec:sqd}.

\subsection{Energy sensitivity and finite-shot propagation}
\label{subsec:scattering_sensitivity}

With the interaction and confinement parameters held fixed, the
finite-shot uncertainty is propagated to first order as
\begin{equation}
\sigma_{\mathcal K}
=\left|\frac{\partial\mathcal K_\omega}{\partial E}
 \right|_{E=E_c}\sigma_E.
\label{eq:app_sigma_K}
\end{equation}
The analytic derivative is given in
Eq.~\eqref{eq:app_derivative_K} of Appendix F. At the benchmark energy it is
\begin{equation}
\left.\frac{\partial\mathcal K_{3.7}}{\partial E}\right|_{E_2}
\simeq0.28942~\mathrm{fm}^{-1}\mathrm{MeV}^{-1}.
\label{eq:scattering_local_sensitivity}
\end{equation}
Thus a small spectral-center displacement produces
\begin{equation}
\Delta\mathcal K_{3.7}
\simeq 2.894\times10^{-4}~\mathrm{fm}^{-1}
       \left(\frac{\Delta E_c}{1~\mathrm{keV}}\right).
\label{eq:scattering_energy_transfer}
\end{equation}
A \(1\keV\) shift corresponds to approximately \(0.2154\%\)
of the exact finite-confinement value at this level. The derivative
therefore supplies a quantitative link between spectral precision and
precision of the scattering input, rather than judging the circuit
only by its accepted-shot yield.

Evaluating the derivative at each fitted center gives propagated
uncertainties of approximately
\(2.01\times10^{-4}~\mathrm{fm}^{-1}\) for IonQ and
\(1.47\times10^{-4}~\mathrm{fm}^{-1}\) for the interleaved IBM
batch. These are conditional finite-shot uncertainties: they retain the
assumptions of the energy fit and do not include batch-to-batch
variation, interaction-model uncertainty, or the uncertainty of a later
confinement extrapolation. Their difference cannot be attributed solely
to ancilla recycling because the processors, native operations, scan
protocols, and effective shot allocations differ.

\subsection{Batch-resolved finite-confinement results}
\label{subsec:scattering_hardware_results}
Applying Eq.~\eqref{eq:MERE_single_level} to the two primary
hardware centers gives
$\mathcal K_{3.7}^{\mathrm{IonQ}} =0.13419\pm0.00020~\mathrm{fm}^{-1}$ and
$\mathcal K_{3.7}^{\mathrm{IBM,I}} =0.13435\pm0.00015~\mathrm{fm}^{-1}$.
Both are consistent with the exact finite-confinement reference within
their quoted uncertainties. The interleaved IBM offset has magnitude
approximately \(4.9\times10^{-6}~\mathrm{fm}^{-1}\), below its
propagated finite-shot uncertainty. The other IBM batches retain the
variation seen in their fitted energy centers: the forward batch maps
to a larger \(\mathcal K_{3.7}\), while the reverse batch maps to
a smaller value. The deterministic conversion does not remove the
repeatability issue identified in Sec.~\ref{subsec:batch_repeatability}.

For a phase-like representation of the same input, define the
finite-confinement diagnostic angle
\begin{equation}
\delta_{0,\omega}^{\rm trap}(E)
=\operatorname{\texttt{atan2}}[k(E),\mathcal K_\omega(E)].
\label{eq:app_trap_angle}
\end{equation}
Here \(\operatorname{\texttt{atan2}}(y,x)=\operatorname{Arg}(x+iy)\),
with the continuous branch convention described in
Appendix~\ref{app:scattering}. Both arguments in
Eq.~\eqref{eq:app_trap_angle} are expressed in
\(\mathrm{fm}^{-1}\). Its uncertainty is propagated from the
same \(E_c\), including the energy dependence of both \(k\)
and \(\mathcal K_\omega\), as specified in
Eq.~\eqref{eq:app_derivative_angle} of Appendix F.

Table~\ref{tab:app_scattering_full_precision} collects the
postprocessing results for all hardware batches. The columns are
transformations of the same fitted energies, not independent QPU
measurements. In particular, the angle column is a finite-trap
diagnostic, not a free-space phase shift.

\begin{table}[!htbp]
\caption{
Finite-confinement scattering postprocessing at \(\omega=3.7\MeV\).
The exact row uses the reference trapped level; hardware rows use the
fixed-time fitted centers of Table~\ref{tab:hardware}. All quoted
uncertainties contain finite-shot propagation from the fitted energy
center only. The angle \(\delta_{0,3.7}^{\rm trap}\) is defined by
Eq.~\eqref{eq:app_trap_angle} and is not a free-space phase shift.
Additional digits are retained for numerical comparison; the displayed
energies and derived quantities are rounded separately.
}
\label{tab:app_scattering_full_precision}
\centering
\small
\begin{adjustbox}{max width=\linewidth}
\begin{tabular}{lccc}
\toprule
Data set & \(E_c\) (MeV) & \(\mathcal K_{3.7}\) (\(\mathrm{fm}^{-1}\))
& \(\delta_{0,3.7}^{\rm trap}\) (deg)\\
\midrule
Exact & 17.134039 & 0.134358 & 78.193565\\
IonQ Forte-1, static
& \(17.133471\pm0.000694\)
& \(0.134193\pm0.000201\)
& \(78.207415\pm0.016922\)\\
IBM Aachen 1, forward
& \(17.135578\pm0.000545\)
& \(0.134803\pm0.000158\)
& \(78.156038\pm0.013289\)\\
IBM Aachen 2, reverse
& \(17.133439\pm0.000529\)
& \(0.134184\pm0.000153\)
& \(78.208196\pm0.012899\)\\
IBM Aachen 3, interleaved
& \(17.134022\pm0.000507\)
& \(0.134353\pm0.000147\)
& \(78.193980\pm0.012363\)\\
\bottomrule
\end{tabular}
\end{adjustbox}
\end{table}

The agreement of the static and interleaved results demonstrates that
the reduced-width implementation preserves the specified scattering
input within its measured finite-shot precision. At the same time,
reporting all three IBM batches makes explicit that stable
postselection attenuation does not establish a stable value of the
scattering input. The conversion therefore carries both the successful
spectral estimate and its repeatability qualification into the nuclear
analysis.

\subsection{Finite confinement and the path to free-space observables}
\label{subsec:scattering_continuum_scope}
The analytic square-well solution allows the effect of finite
confinement to be distinguished from a hardware center error.
At the exact target energy,
$\delta_{0,3.7}^{\rm trap}(E_2) =78.19357^\circ$ and
$\delta_0^{\rm SW}(E_2) =78.14319^\circ$,
with \(k(E_2)\cot\delta_0^{\rm SW}(E_2)
\simeq0.13495~\mathrm{fm}^{-1}\).
The difference of approximately \(0.0504^\circ\) is present even when
the exact trapped energy is used as input.
It is therefore a finite-confinement
effect in this benchmark, not a QPU error or a contribution to the
finite-shot error bar. Reducing the statistical uncertainty on one
trapped level does not remove this difference.

A free-space reconstruction requires levels at several trap strengths,
followed by interpolation in energy and extrapolation to vanishing
confinement \cite{Luu2010,Zhang2020,Wang2024NuclearScattering}.
To distinguish this reconstruction from the trap expression in
Eq.~\eqref{eq:MERE_single_level}, let \(\mathcal F_\omega(E)\)
denote the smooth effective-range function fitted to the samples
\(\{E_n(\omega),\mathcal K_\omega[E_n(\omega)]\}\).
In the analytic weak-confinement expansion at fixed energy,
only even powers of \(\omega\) enter \cite{Zhang2020}, giving
\begin{equation}
\begin{aligned}
\mathcal F_\omega(E)
&=\mathcal K_0(E)+\mathcal K_1(E)\omega^2+\mathcal O(\omega^4),\\
\mathcal K_0(E)&=k(E)\cot\delta_0(E).
\end{aligned}
\label{eq:app_trap_extrapolation}
\end{equation}
The extrapolation acts on the fitted spectral data, not on the
Gamma-function ratio at an arbitrary fixed energy without reference
to the trapped spectrum. The single level studied here supplies one
sample and does not determine the functions in
Eq.~\eqref{eq:app_trap_extrapolation}. For the scattering length,
the additional threshold expansion is
\begin{equation}
k\cot\delta_0(k)
=-\frac{1}{a_s}+\frac12 r_e k^2+\mathcal O(k^4),
\qquad
a_s=-\frac{1}{\mathcal K_0(0)},
\label{eq:app_ere}
\end{equation}
where \(r_e\) is the effective range. Neither \(a_s\) nor
\(r_e\) is inferred from the present hardware dataset.

The contribution of this benchmark is the tested spectral step feeding
that reconstruction: the static and recycled-ancilla circuits are
compared on the same trapped Hamiltonian, their center uncertainties
are carried into a defined scattering input, and batch variation is
kept distinct from finite-shot precision and finite-confinement effects.
Repeating this step for additional levels and trap strengths reuses the
one-ancilla structure without increasing its simultaneous ancilla
requirement. The associated spectral, support, and repeatability checks
provide the basis for assessing those additional inputs before they are
combined in a continuum analysis.

\FloatBarrier
\section{Conclusions}
\label{sec:conclusions}

We have realized the same \(R=10\) trapped-nuclear RODEO filter as
a static 12-active-qubit circuit on IonQ Forte-1 and an ancilla-recycled
dynamic three-qubit circuit on IBM Aachen. The ideal all-zero accepted
branch is unchanged, while the active width decreases from
\(n_s+R\) to \(n_s+1\). The interleaved IBM scan gives
\(\Delta E_c=-0.017\pm0.507\keV\), compared with
\(-0.568\pm0.694\keV\) for IonQ; both are consistent with the
exact trapped level within their finite-shot bootstrap uncertainties.
The reduced-width implementation therefore retains a trapped-energy
estimate with comparable finite-shot precision while using 75\% fewer
simultaneously active qubits.

The hardware measurements also distinguish spectral precision from
repeatability. Three independent IBM batches have heterogeneous centers
under their individual finite-shot uncertainties
(\(Q=8.45\), \(p=0.015\)), even though their postselection
amplitudes are comparatively stable. Line-shape comparisons and repeated
central queries probe within-batch sensitivity, whereas separate batches
expose variation not captured by the shot-only error bars.
Calibration-derived Aer simulations do not consistently reproduce the
measured centers and attenuation. These results identify a practical
requirement for repeated spectral estimation: accepted-event yield,
center accuracy, and batch consistency must be assessed separately.
The present data diagnose the variation without separating
submission-order effects from changes between hardware batches.

All central hardware records retain the complete four-configuration OLS
support and consequently reproduce the exact effective-space level by
SQD. This recovery verifies sampled-basis completeness, rather than
correcting the fitted center or reconstructing an exact quantum state.
The larger-space simulator tests show why the distinction matters:
small oscillator-space truncation errors and large observed support
fractions can coexist with substantial sampled-subspace energy errors.
Frequency-retention and Hamiltonian-guided augmentation tests therefore
complement the spectral checks by examining sensitivity to omitted
configurations. They establish distinct roles for basis convergence,
sampled support, and target-branch identification when extending the
nuclear register.

The nuclear significance of the hardware result is quantified by
propagating the fitted energies into the finite-confinement MERE input.
At \(\omega=3.7\MeV\), the static IonQ and interleaved IBM
centers give \(\mathcal K_{3.7}^{\rm IonQ}
=0.13419\pm0.00020~\mathrm{fm}^{-1}\) and
\(\mathcal K_{3.7}^{\rm IBM,I}
=0.13435\pm0.00015~\mathrm{fm}^{-1}\), respectively, consistent
with the exact value \(0.13436~\mathrm{fm}^{-1}\). Their quoted
errors contain finite-shot propagation only. The local sensitivity,
approximately \(2.894\times10^{-4}~\mathrm{fm}^{-1}\) per
\(1\keV\) center shift, connects spectral errors to the quantity
entering the scattering analysis. This validates one finite-confinement
input; free-space phase shifts require additional levels and trap
strengths, with confinement effects and batch variation assessed
separately from statistical precision.

The exactly known four-dimensional reference provides a controlled
test of this spectral step. Ancilla recycling remains a space--time
tradeoff: all ten controlled evolutions are retained, while mid-circuit
measurement and reset introduce latency and additional system-register
exposure. Its practical gain is to remove the cycle-dependent ancilla
allocation, leaving more qubits available for the nuclear register at
fixed total width. The sequential controlled-evolution, measurement,
reset, and recorded-outcome postselection structure can be reused for
other trapped Hamiltonians, with problem-dependent evolutions recompiled
and the resulting spectra validated anew.

The contribution of this work is thus a hardware-tested implementation
together with a validation protocol connecting circuit resources,
measured spectra, sampled support, and a defined nuclear-scattering
input. This three-qubit implementation already provides the starting
point for our subsequent factorized geometric-time RODEO study, which
extends the calculation to four levels at five confinement strengths
\cite{Mun2026FGRodeo}. That study addresses further compression of the
evolution sequence and free-space reconstruction; the present work
establishes the cross-platform benchmark and batch-resolved spectral
tests on which the extension builds. Together, the reusable circuit and
its validation procedure provide a basis for assembling nuclear
scattering inputs from repeated quantum-hardware measurements.

\FloatBarrier

\section*{Acknowledgments}
M.-H.M. was supported by the National Research Foundation of Korea
(NRF) through grants funded by the Ministry of Science and ICT (MSIT)
of the Korean government under Grant Nos. RS-2026-25487837 and
RS-2018-NR031074.
M.-K.C. was supported by the NRF Basic Science Research Program under
Grant Nos. RS-2021-NR060129, RS-2024-00460031, and RS-2025-16071941.
J.-B.P. was supported by NRF grants funded by MSIT and the Ministry
of Education under Grant Nos. RS-2025-24533596 and RS-2025-25400847,
respectively.
E.-J.H. was supported by the NRF under Grant No. RS-2025-00513410.
This work was supported by quantum computing cloud resources provided
by the Korea Institute of Science and Technology Information (KISTI)
through an NRF grant funded by MSIT (Grant No. RS-2025-24534879).
This research was also supported by the ``Quantum Information Science
R\&D Ecosystem Creation'' program through an NRF grant funded by MSIT
(Grant No. 2020M3H3A1110365).

\section*{Data Availability}
The raw QPU counts, QPU job identifiers, circuit files, processed data,
and analysis scripts supporting the findings of this article are available
from the corresponding authors upon reasonable request.

\FloatBarrier
\appendix
\numberwithin{equation}{section}
\numberwithin{table}{section}
\numberwithin{figure}{section}
\renewcommand{\theequation}{\thesection\arabic{equation}}
\renewcommand{\thetable}{\thesection\arabic{table}}
\renewcommand{\thefigure}{\thesection\arabic{figure}}
\renewcommand{\theHequation}{appendix.\arabic{section}.\arabic{equation}}
\renewcommand{\theHtable}{appendix.\arabic{section}.\arabic{table}}
\renewcommand{\theHfigure}{appendix.\arabic{section}.\arabic{figure}}

\section*{Appendices}
These appendices provide the numerical definitions and supporting
methods used in the main text. Appendix~\ref{app:benchmark} specifies
the OLS Hamiltonian and reference spectrum;
Appendices~\ref{app:equivalence} and \ref{app:hardware} record the
filter construction, interrogation times, scan protocols, compiled-circuit
metrics, and raw hardware counts. Appendix~\ref{app:fitting} gives
the line-shape models and bootstrap procedure;
Appendix~\ref{app:diagnostics} gives the calibration-derived spectra,
frequency-ranked subspace tests, larger-space completeness diagnostics,
and full-space Ritz residual; and
Appendix~\ref{app:scattering} gives the analytic square-well reference
and uncertainty-propagation derivatives. The main text presents the hardware spectral tests, observed-support
audit, and batch-resolved scattering inputs in
Secs.~\ref{sec:hardware}--\ref{sec:nuclear_scattering_implications}.

\FloatBarrier
\section{OLS benchmark Hamiltonian and target state}
\label{app:benchmark}

The benchmark is the four-dimensional Okubo--Lee--Suzuki (OLS)
effective Hamiltonian for the model $0^+$ square-well $np$ system in a
harmonic trap with $\omega=3.7\MeV$.  The computational-basis ordering is
$\{|00\rangle,|01\rangle,|10\rangle,|11\rangle\}$.  These bit strings
label OLS model-space basis vectors and are not neutron or proton
occupation labels.

\begin{equation}
\Heff=
\begin{pmatrix}
10.917568 &  5.720650 & -0.891124 &  1.085683\\
 5.720650 & 30.478581 & -1.902297 & -1.382753\\
-0.891124 & -1.902297 & 24.852377 & -1.868274\\
 1.085683 & -1.382753 & -1.868274 & 17.533252
\end{pmatrix}\MeV .
\label{eq:app_hamiltonian}
\end{equation}

The target of the hardware scans is
\begin{equation}
\eref=E_2=17.134039\MeV .
\label{eq:app_target_energy}
\end{equation}
It is a discretized positive-energy state of the relative motion, not a
bound-state energy.  The hardware scans use
$|\psi_0\rangle=|11\rangle$, whose squared overlap with the target OLS
eigenstate is $0.8941925$.
The OLS eigenenergies and computational-basis probabilities are
summarized in Table~\ref{tab:app_eigensystem}. The precision displayed for the matrix
entries is lower than that of the tabulated eigenenergies.

\begin{table}[!htbp]
\centering
\caption{OLS eigenenergies and computational-basis probabilities at
$\omega=3.7\MeV$. The last four columns give
$|\langle z|E_j\rangle|^2$ for the indicated basis state $z$.}
\label{tab:app_eigensystem}

\small
\begin{adjustbox}{max width=\linewidth}
\begin{tabular}{lccccc}
\toprule
State & $E$ (MeV) & $z=00$ & $z=01$ & $z=10$ & $z=11$\\
\midrule
$E_1$ &  9.1295508 & 0.9022145 & 0.0702677 & 0.0000042 & 0.0275136\\
$E_2$ & 17.1340393 & 0.0326833 & 0.0033792 & 0.0697450 & 0.8941925\\
$E_3$ & 24.9160822 & 0.0002227 & 0.0548908 & 0.8679143 & 0.0769722\\
$E_4$ & 32.6021052 & 0.0648795 & 0.8714623 & 0.0623365 & 0.0013217\\
\bottomrule
\end{tabular}
\end{adjustbox}

\end{table}

\FloatBarrier
\section{Static--dynamic equivalence and scan design}
\label{app:equivalence}

At query energy $E$, RODEO cycle $j$ applies
\begin{equation}
U_j(E)=\exp[-i(\Heff-EI)t_j].
\label{eq:app_controlled_evolution}
\end{equation}
After the standard Hadamard--controlled-$U_j$--Hadamard sequence,
postselection of the ancilla in $|0\rangle$ gives the conditional system
operator
\begin{equation}
M_j(E)=\frac{I+U_j(E)}{2}.
\label{eq:app_kraus}
\end{equation}
The accepted state and all-zero success probability after $R$ cycles are
\begin{equation}
\begin{aligned}
|\widetilde\psi_R(E)\rangle
&=M_R(E)\cdots M_1(E)|\psi_0\rangle,\\
P_0(E)&=\|\widetilde\psi_R(E)\|^2.
\end{aligned}
\label{eq:app_accepted_state}
\end{equation}

In the static circuit, the system interacts sequentially with $R$
different ancillas and all ancillas are measured at the end.  In the
dynamic circuit, the same ancilla is measured after each cycle, its
outcome is stored, and it is reset before reuse.  Conditioning on the
same all-zero record therefore produces the same ideal accepted branch.
This equivalence does not imply identical hardware noise: the two
realizations differ in measurement timing, reset, idle exposure, native
compilation, and processor architecture.

For $|\psi_0\rangle=\sum_\lambda c_\lambda|\lambda\rangle$,
\begin{equation}
P_{\bm t}(E)=
\sum_\lambda |c_\lambda|^2
\prod_{j=1}^{R}
\cos^2\!\left[\frac{(E-E_\lambda)t_j}{2}\right].
\label{eq:app_fixed_response}
\end{equation}
The active widths are
\begin{equation}
W_{\rm stat}=n_s+R,
\qquad
W_{\rm dyn}=n_s+1.
\label{eq:app_widths}
\end{equation}
For $n_s=2$ and $R=10$, the width is reduced from 12 to 3 while all ten
controlled evolutions are retained.

The interrogation times are deterministic symmetric Gaussian quantiles,
\begin{equation}
t_j=\sigma\,\Phi^{-1}\!\left(\frac{j-1/2}{R}\right),
\qquad
\sigma=21\MeV^{-1},
\label{eq:app_times}
\end{equation}
where $\Phi^{-1}$ is the standard-normal quantile function.
The resulting numerical interrogation times are listed in
Table~\ref{tab:app_times}.

\begin{table}[!htbp]
\caption{Interrogation times used in every hardware and simulator scan.}
\label{tab:app_times}
\centering

\small
\begin{adjustbox}{max width=\linewidth}
\begin{tabular}{cccc}
\toprule
$j$ & $t_j$ (MeV$^{-1}$) & $j$ & $t_j$ (MeV$^{-1}$) \\
\midrule
1 & $-34.541926$ & 6  & $+2.638888$ \\
2 & $-21.765101$ & 7  & $+8.091730$ \\
3 & $-14.164285$ & 8  & $+14.164285$ \\
4 & $-8.0917298$ & 9  & $+21.765101$ \\
5 & $-2.6388883$ & 10 & $+34.541926$\\
\bottomrule
\end{tabular}
\end{adjustbox}

\end{table}

The seven physical offsets are
\begin{equation}
E-E_2=(-40,-28,-16,0,+16,+28,+40)\keV .
\label{eq:app_offsets}
\end{equation}
IonQ and IBM Run~1 use this ascending order; IBM Run~2 uses the reverse
order.  IBM Run~3 uses
\begin{equation}
(0,+16,-16,+28,-28,+40,-40,0)\keV,
\label{eq:app_interleaved_order}
\end{equation}
so the central circuit is executed at the beginning and end of the job.
Each circuit contains 500 shots.  The two Run~3 central records are kept
separate for the within-job check and pooled only for the primary
seven-point fit.

\FloatBarrier
\section{Hardware execution, compiled circuits, and raw counts}
\label{app:hardware}

All primary QPU results use raw counts.  Debiasing, twirling, dynamical
decoupling, and error mitigation are disabled.  The three IBM batches
use the same archived ISA circuits, rearranged only in submission order.
The hardware job identifiers and scan protocols are listed in
Table~\ref{tab:app_jobs}.

\begin{table}[!htbp]
\centering
\caption{Hardware jobs and scan protocols.}
\label{tab:app_jobs}

\small
\begin{adjustbox}{max width=\linewidth}
\begin{tabular}{lccl}
\toprule
Platform/batch & Job identifier & Circuits $\times$ shots & Protocol\\
\midrule
IonQ Forte-1 & \texttt{019f3c67-ef8c-7717-8620-cae2e9bb4cbf} & $7\times500$ & ascending\\
IBM Aachen 1 & \texttt{d9tb4jhnf51c73e8c20g} & $7\times500$ & forward\\
IBM Aachen 2 & \texttt{d9td6ipnf51c73e8dsgg} & $7\times500$ & reverse\\
IBM Aachen 3 & \texttt{d9tukjucseuc73f2e1d0} & $8\times500$ & interleaved\\
\bottomrule
\end{tabular}
\end{adjustbox}

\end{table}

The IBM circuits act on physical qubits 153--155.  Although the backend
contains 156 qubits, only these three appear in the submitted circuits.
Each query circuit contains 112 CZ gates, ten mid-circuit ancilla
measurements, nine resets, and two terminal system measurements.
The query-dependent compiled-circuit metrics are summarized in
Table~\ref{tab:app_metrics}.

\begin{table}[!htbp]
\caption{IBM Aachen ISA-circuit metrics.}
\label{tab:app_metrics}
\centering

\small
\begin{adjustbox}{max width=\linewidth}
\begin{tabular}{ccccc}
\toprule
Offset (keV) & Depth & Instructions & CZ & Reset\\
\midrule
$-40$ & 450 & 633 & 112 & 9\\
$-28$ & 450 & 633 & 112 & 9\\
$-16$ & 449 & 627 & 112 & 9\\
$0$   & 450 & 627 & 112 & 9\\
$+16$ & 453 & 638 & 112 & 9\\
$+28$ & 449 & 632 & 112 & 9\\
$+40$ & 447 & 632 & 112 & 9\\
\bottomrule
\end{tabular}
\end{adjustbox}

\end{table}

For $N_{\rm sh}$ shots, the measured estimator is
\begin{equation}
\widehat P_0(E)=\frac{N_0(E)}{N_{\rm sh}},
\label{eq:app_success_estimator}
\end{equation}
where $N_0(E)$ is the number of all-zero ancilla records.  The raw
success counts are listed in Table~\ref{tab:app_rawcounts}.

The Run~3 central entry combines $325/500$ at the beginning and $335/500$
at the end, giving $660/1000=0.660$.

\begin{table}[!htbp]
\centering
\caption{Raw all-zero-ancilla success counts.
Each entry contains 500 shots, except the pooled IBM Run~3 center,
which contains 1000.}
\label{tab:app_rawcounts}

\small
\begin{adjustbox}{max width=\linewidth}
\begin{tabular}{lrrrrrrr}
\toprule
Data set & $-40$ & $-28$ & $-16$ & $0$ & $+16$ & $+28$ & $+40$\\
\midrule
IonQ Forte-1 & 67 & 126 & 177 & 244 & 188 & 117 & 50\\
IBM Run 1 & 55 & 125 & 231 & 313 & 268 & 151 & 63\\
IBM Run 2 & 76 & 148 & 272 & 347 & 253 & 148 & 65\\
IBM Run 3 & 57 & 147 & 255 & 660 & 264 & 137 & 59\\
\bottomrule
\end{tabular}
\end{adjustbox}

\end{table}

\FloatBarrier
\section{Peak fits and hardware repeatability}
\label{app:fitting}

This appendix records the line-shape definitions and resampling
procedure used for the hardware analysis of Sec.~\ref{sec:hardware}.
The numerical fit comparison and internal-control figure are presented
in the main text as Table~\ref{tab:app_fitmodels} and
Fig.~\ref{fig:app_fitcontrols}, respectively.

Define
\begin{equation}
f_{\bm t}(x)=\prod_{j=1}^{R}\cos^2(xt_j/2).
\label{eq:app_fixed_filter}
\end{equation}
The primary local model is
\begin{equation}
P_{\rm fit}^{(t)}(E;E_c,K)=Kf_{\bm t}(E-E_c),
\label{eq:app_fixed_fit}
\end{equation}
and the Gaussian-averaged cross-check is
\begin{equation}
P_{\rm fit}^{(G)}(E;E_c,K)
=K\left[\frac{1+e^{-\sigma^2(E-E_c)^2/2}}{2}\right]^R.
\label{eq:app_gaussian_fit}
\end{equation}
The fits use binomial standard errors and two free parameters,
\((E_c,K)\). The common noiseless amplitude reference is
\begin{equation}
K_{\rm loc}^{\rm ex}\simeq0.894.
\label{eq:app_amplitude_reference}
\end{equation}
It is obtained by fitting the noiseless fixed-time response at the same
seven query energies and is therefore not identical to the isolated
target-state overlap \(\lvert\langle E_2|\psi_0\rangle\rvert^2\).

For each parametric-bootstrap replica, a success count is drawn
independently at every query 
%from the fitted binomial probability and the original shot number. 
from a binomial distribution with the observed success fraction and the original shot number.
The replica is refitted with the same model.
The reported \(\sigma_E\) is the standard deviation of 
%3000 successful bootstrap centers. 
2000 successful bootstrap centers for IonQ and 3000 for IBM.
For the primary Run~3 fit, the pooled
central count has 1000 shots and each other point has 500 shots.
These resamples describe %finite-shot fluctuations about the fitted response; 
finite-shot fluctuations about the observed success fractions; 
they do not sample the changes between independent hardware
batches.

The common-center consistency diagnostic is defined in
Eq.~\eqref{eq:app_cochran} and evaluated in
Sec.~\ref{subsec:batch_repeatability}. The two Run~3 central records are analyzed separately in the
repeated-center comparison of Sec.~\ref{subsec:hardware_controls}. The additional
within-job test fits all eight raw query records with a binomial model
including a linear submission-position term and compares it with the
model without that term using a likelihood-ratio test. Its reported
\(p=0.969\), together with the central-record sensitivity checks,
is discussed in Sec.~\ref{subsec:hardware_controls}.

\FloatBarrier

\section{Noisy-simulator and sampled-subspace diagnostics}
\label{app:diagnostics}

This appendix records the detailed sensitivity tests supporting the
sampled-support audit in Sec.~\ref{sec:sqd}. The observed hardware
counts remain in Table~\ref{tab:app_systemcounts}; the frequency-ranked
subspaces, larger-space tests, and residual identities are given here.

\subsection{Calibration-derived spectra}
\label{subsec:app_simulator_records}

% AUTHOR CHECK BEFORE SUBMISSION: document the calibration snapshots,
% noise-model construction (including reset/mid-circuit measurement),
% simulator/version, and shot counts, or cite an archived analysis script.
Calibration-derived Aer simulations associated with the first two IBM
campaigns give
\begin{equation}
\begin{aligned}
\Delta E_c^{\rm Aer,1}&=+0.015\pm0.475\keV,\\
\Delta E_c^{\rm Aer,2}&=-0.845\pm0.488\keV.
\end{aligned}
\label{eq:app_aer_centers}
\end{equation}
Their amplitude ratios are 0.850 and 0.848, compared with 0.703
and 0.761 for the corresponding hardware batches. The comparison
with the measured spectra is discussed in
Sec.~\ref{subsec:hardware_aer}.

\subsection{Frequency-ranked hardware subspaces}
\label{subsec:sqd_frequency_sensitivity}

\begin{figure}[!htbp]
\centering
\includegraphics[width=0.92\textwidth]{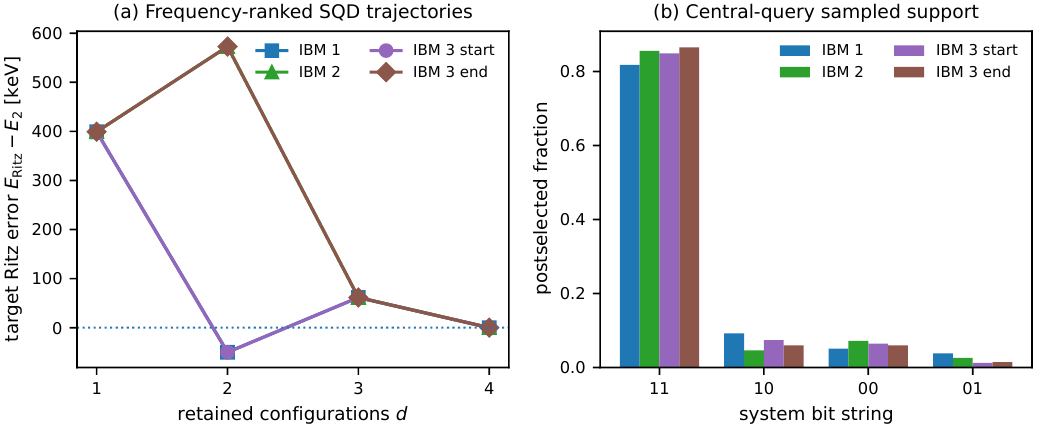}
\caption{
Sampled-support diagnostics from the IBM central-query records.
(a) Target-related Ritz energies as configurations are retained in
frequency-ranked order. The error is measured relative to the exact
OLS target \(E_2\), not to the fitted hardware center.
(b) Postselected configuration fractions. All records contain the
complete four-configuration OLS space, while the intermediate Ritz
values depend on which configuration enters first. The two Run~3
records come from the same hardware job. For the tied end record, the
displayed path places \(00\) before \(10\); both choices have
the same observed count.
}
\label{fig:app_sqd}
\end{figure}

The hardware configuration counts are listed in Table~\ref{tab:app_systemcounts}.
To examine the effect of
retaining only the most frequent configurations, the sampled strings
are ordered by their observed counts and the Hamiltonian is
diagonalized in the resulting nested subspaces.
Figure~\ref{fig:app_sqd}(a) shows the target-related Ritz trajectories
for the IBM central records, while panel (b) shows their configuration
fractions. All trajectories recover the exact level at dimension four,
but their intermediate values depend on the configurations retained.

For example, the two-dimensional subspaces give

$E_{\{11,10\}}-E_2 \simeq-50.1\keV$ and
$E_{\{11,00\}}-E_2 \simeq+572.8\keV$.

Each energy denotes the target-related Ritz root in the indicated
subspace. The Run~3 end record makes this sensitivity particularly
explicit: \(10\) and \(00\) are tied at 20 counts, so frequency
alone does not choose between these two subspaces. The displayed
trajectory follows the recorded tie choice; changing that choice alters
the intermediate energy but not the complete-support endpoint.
These are alternative subspace analyses of the same data, rather than
additional measurements or a statistical uncertainty interval.

The errors
are much larger than the finite-shot center uncertainties in
Table~\ref{tab:hardware}. They do not contradict those uncertainties:
they arise from removing basis vectors before classical diagonalization,
not from refitting the measured success probabilities. This example
shows why the observed frequency of a configuration cannot by itself
specify its relevance to a target eigenvalue. The projected energy also
depends on its Hamiltonian couplings and on the eigenstate branch being
followed.

\subsection{Larger-space tests and separation of truncation errors}
\label{subsec:sqd_larger_spaces}

% AUTHOR CHECK BEFORE SUBMISSION: document the oscillator-basis frequency,
% precise N/reference-space definition, target branch, state preparation,
% sampling shot count, and noise model used to generate the large-space
% samples. These values are not specified in the supplied manuscript.
The larger-space calculations are simulator-side stress tests and are
not additional QPU executions. Unlike the four-dimensional OLS
Hamiltonian used on hardware, these tests use direct oscillator-basis
truncations
\begin{equation}
H_N=P_N H_{\rm bare}P_N,
\label{eq:sqd_direct_truncation}
\end{equation}
represented within the \(N\)-dimensional space selected by
\(P_N\). Their target energies \(E_N\) are compared with an
\(N=300\) reference. For an observed subset \(S_{\rm obs}\),
\(E_{S_{\rm obs}}\) is the Ritz root selected by its overlap with
the corresponding exact finite-\(N\) target state. This reference
makes branch assignment possible in the benchmark; it is not an
assumption that an unknown target state is available in a production
calculation.

Two distinct errors must be separated: the finite oscillator-space
truncation and the additional projection onto the sampled subspace.
Their relation can be written as
\begin{equation}
E_{S_{\rm obs}}-E_{300}
=\bigl(E_N-E_{300}\bigr)
 +\bigl(E_{S_{\rm obs}}-E_N\bigr).
\label{eq:sqd_error_decomposition}
\end{equation}
Table~\ref{tab:app_larger} reports both contributions. The
\(N=300\) calculation is a numerical reference, not an assertion
that finite-basis effects vanish exactly at that dimension.

\begin{table}[!htbp]
\caption{
Representative larger-space sampled-subspace diagnostics from finite-shot
noisy-simulator realizations. The first energy difference measures
oscillator-space truncation relative to the \(N=300\) reference;
the last measures the additional sampled-subspace error relative to the
complete finite-\(N\) target. Each row is one realization, not an
ensemble average or a hardware result. Signed errors refer to the
overlap-tracked target branch.
}
\label{tab:app_larger}
\centering
\small
\begin{adjustbox}{max width=\linewidth}
\begin{tabular}{rrcc}
\toprule

\(N\) & \(E_N-E_{300}\) (keV) &
\(\lvert S_{\rm obs}\rvert/N\) &
\(E_{S_{\rm obs}}-E_N\) (keV) \\
\midrule
16  & \(+16807.177\) & \(16/16\) & \(<10^{-6}\) \\
32  & \(+3067.718\)  & \(32/32\) & \(<10^{-6}\) \\
64  & \(+47.134\)    & \(63/64\) & \(+1537.4\) \\
128 & \(+0.022\)     & \(117/128\) & \(-1091.0\) \\
\bottomrule
\end{tabular}
\end{adjustbox}
\end{table}

For \(N=16\) and 32, complete sampled support recovers the
finite-\(N\) level, while the oscillator-space truncation relative
to \(E_{300}\) remains substantial. Conversely, for \(N=128\),
the complete-space target differs from the reference by only
\(0.022\keV\), but the observed \(117/128\) subspace leaves
an error of \(-1091.0\keV\). Even observing 63 of 64
configurations leaves \(+1537.4\keV\) in the displayed
\(N=64\) example. Thus convergence of the underlying basis and
coverage of the sampled basis are separate requirements; neither a
small \(E_N-E_{300}\) nor a large observed fraction controls the
other error.

The roots are matched to an excited target state by overlap rather than
by a fixed ordered eigenvalue index. A matched Ritz root may therefore
lie below the target energy without violating the variational bound for
ordered Ritz eigenvalues.
Because each row represents one finite-shot realization, the table
illustrates possible failure modes rather than their occurrence
probabilities or an asymptotic scaling law.

\subsection{Hamiltonian-aware completeness checks}
\label{subsec:sqd_completeness_checks}

\begin{figure}[!htbp]
\centering
\includegraphics[width=0.92\textwidth]{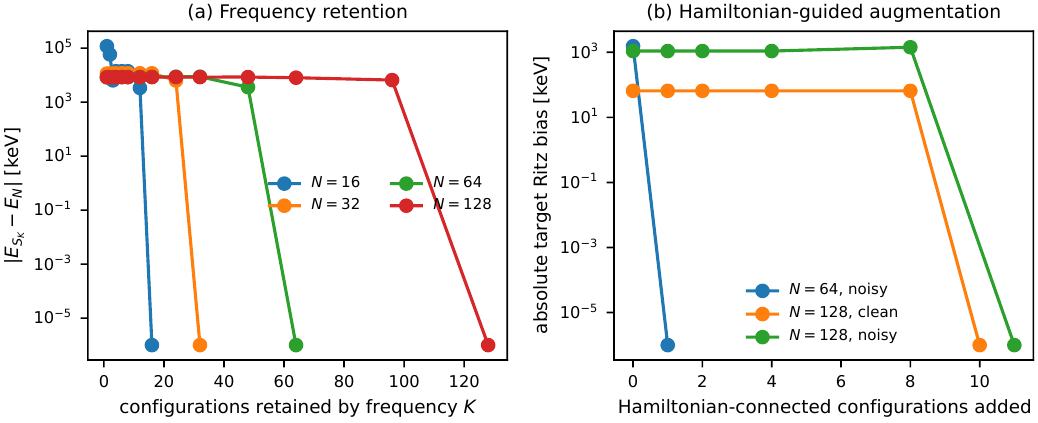}
\caption{
Larger-space completeness tests using simulator-generated subspaces.
(a) Absolute target-branch Ritz error as configurations are retained by
sampled frequency. The horizontal-axis symbol \(K\) counts retained
configurations and is unrelated to the fitted probability amplitude in
Sec.~\ref{sec:compression}.
(b) Representative Hamiltonian-guided augmentation for \(N=64\)
and 128. Exact zeros are displayed at \(10^{-6}\keV\).
Here ``clean'' and ``noisy'' denote subspaces obtained from noiseless-
and noisy-simulator samples, respectively. These trajectories describe
completeness sensitivity for the displayed samples, not confidence
intervals or a comparison of QPU performance.
}
\label{fig:app_larger}
\end{figure}

Figure~\ref{fig:app_larger}(a) shows the absolute target-branch error
as configurations are retained by frequency. Long intervals with large
errors can persist before the necessary configurations enter the
subspace. Missing configurations can instead be examined through their
couplings to the retained set. The boundary-coupling score used in the
augmentation tests is
\begin{equation}
b_j(S)=\sum_{i\in S}\lvert(H_N)_{ji}\rvert,
\qquad j\notin S,
\label{eq:app_boundary_score}
\end{equation}
and configurations ranked by this score are added before
rediagonalization. This score uses Hamiltonian information absent from
a frequency-only ranking.

The augmentation trajectories in Fig.~\ref{fig:app_larger}(b) show
how restoring omitted configurations changes the target-related
energy. For \(N=64\), adding the single missing configuration
completes the space and recovers its exact finite-\(N\) eigenvalue.
The \(N=128\) trajectories also show that several additions can
leave a large error before a marked change occurs. These are
completeness sensitivities, not statistical confidence intervals; they
do not establish a general sparse-recovery rate or an advantage over
all alternative selection rules.

\subsection{Projected and full-space Ritz residuals}
\label{subsec:app_ritz_residual}

The following algebra clarifies 
%the residual diagnostic mentioned in Appendix~\ref{subsec:sqd_completeness_checks}; 
the full-space Ritz residual diagnostic; 
it is not an additional hardware or simulator dataset. Let \(H\) denote either \(\Heff\)
or a complete finite-\(N\) Hamiltonian \(H_N\), and let
\(V_S\) contain orthonormal computational-basis columns. Define
\begin{equation}
H_S=V_S^\dagger H V_S,
\qquad \Pi_S=V_SV_S^\dagger.
\label{eq:app_sqd_projection}
\end{equation}
For a normalized projected eigenvector \(u_S\) satisfying
\(H_Su_S=E_Su_S\), the embedded Ritz vector is
\(\lvert\phi_S\rangle=V_Su_S\). Its full-space residual obeys
\begin{equation}
\begin{aligned}
\lvert r_S\rangle
&=(H-E_S I)\lvert\phi_S\rangle
 =(I-\Pi_S)H\lvert\phi_S\rangle,\\
V_S^\dagger\lvert r_S\rangle&=0.
\end{aligned}
\label{eq:app_full_ritz_residual}
\end{equation}
Thus solving the projected eigenproblem accurately does not force the
full-space residual to vanish. For computational-basis selection,
\begin{equation}
\lVert r_S\rVert^2
=\sum_{j\notin S}
 \left|\sum_{i\in S}H_{ji}(\phi_S)_i\right|^2.
\label{eq:app_residual_components}
\end{equation}
This expression contains the amplitudes and their interference,
whereas \(b_j(S)\) is a sum of absolute matrix elements. The
boundary score is therefore not itself a residual norm or an energy
error bound. Complete support makes the residual vanish in the chosen
finite space. It does not establish convergence of that space to the
untruncated problem, nor does a small residual alone assign an
excited-state branch. These distinctions motivate separate basis,
support, and branch checks; no new residual values or residual-derived
uncertainties are reported here.

\FloatBarrier
\section{Nuclear-scattering postprocessing}
\label{app:scattering}

This appendix gives the analytic square-well reference and derivatives
used in Sec.~\ref{sec:nuclear_scattering_implications}. The
finite-confinement conversion, propagated-error definition, and
batch-resolved numerical results are presented in the main text in
Eqs.~\eqref{eq:MERE_single_level} and \eqref{eq:app_sigma_K}
and Table~\ref{tab:app_scattering_full_precision}.

\subsection{Square-well matching and the phase convention}
\label{subsec:app_square_well_reference}

The square-well parameters are \(V_0=48.0002\MeV\),
\(R_0=1.70134~\mathrm{fm}\), and \(\mu=469.460\MeV\).
Using \(\hbar c=197.326980~\mathrm{MeV\,fm}\), define
\begin{equation}
q_B=\frac{\sqrt{2\mu(V_0-B)}}{\hbar c},\qquad
\alpha_B=\frac{\sqrt{2\mu B}}{\hbar c}.
\label{eq:app_bound_momenta}
\end{equation}
The bound-state matching condition
\begin{equation}
q_B\cot(q_BR_0)=-\alpha_B
\label{eq:app_bound_matching}
\end{equation}
gives \(B=2.22002\MeV\). Defining
\(\kappa_0=\sqrt{2\mu V_0}/(\hbar c)\), the exact square-well
scattering length is
\begin{equation}
a_s=R_0\left[1-\frac{\tan(\kappa_0R_0)}{\kappa_0R_0}\right]
=5.20134~\mathrm{fm}.
\label{eq:app_scattering_length}
\end{equation}
These are properties of the input interaction, not quantities extracted
from the single QPU level.

For positive energy, define
\begin{equation}
k=\frac{\sqrt{2\mu E}}{\hbar c},\qquad
q=\frac{\sqrt{2\mu(E+V_0)}}{\hbar c}.
\label{eq:app_scattering_momenta}
\end{equation}
Matching \(u_{\rm in}=A\sin(qr)\) and
\(u_{\rm out}=B'\sin[kr+\delta_0^{\rm SW}(E)]\) at \(R_0\)
gives
\begin{equation}
q\cot(qR_0)=k\cot[kR_0+\delta_0^{\rm SW}(E)],
\label{eq:app_scattering_matching}
\end{equation}
and the continuous physical branch may be evaluated as
\begin{equation}
\delta_0^{\rm SW}(E)=-kR_0+\operatorname{atan2}
\!\left[k\sin(qR_0),q\cos(qR_0)\right]+n\pi.
\label{eq:app_square_well_phase}
\end{equation}
We define \(\operatorname{atan2}(y,x)=\operatorname{Arg}(x+iy)\)
with principal value in \((-\pi,\pi]\).
The integer \(n\)
selects the continuous physical branch approaching \(\pi\) as
\(E\to0^+\), because the well supports one \(S\)-wave bound
state. The same quadrant convention and continuous \(\pi\)-branch
choice are used for the finite-trap angle in
Eq.~\eqref{eq:app_trap_angle}; this does not identify that diagnostic
with the free-space square-well phase shift.

\subsection{Derivatives for finite-shot propagation}
\label{subsec:app_scattering_derivatives}

All derivatives below hold at fixed \(\omega\) and \(\mu\).
For the unit convention of Eq.~\eqref{eq:MERE_single_level},
\begin{equation}
\frac{\partial\mathcal K_\omega}{\partial E}
=\frac{\mathcal K_\omega(E)}{2\omega}
\left[\psi\!\left(\frac14-\frac{E}{2\omega}\right)
-\psi\!\left(\frac34-\frac{E}{2\omega}\right)\right],
\label{eq:app_derivative_K}
\end{equation}
where \(\psi(z)=\Gamma'(z)/\Gamma(z)\) is the digamma function.
At the target level this gives approximately
\(0.28942~\mathrm{fm}^{-1}\mathrm{MeV}^{-1}\), as reported in
Eq.~\eqref{eq:scattering_local_sensitivity}. Applying
Eq.~\eqref{eq:app_sigma_K} requires \(\sigma_E\) in MeV when
this derivative is used.

The derivative of the finite-confinement diagnostic angle is
\begin{equation}
\begin{aligned}
\frac{d\delta_{0,\omega}^{\rm trap}}{dE}
&=\frac{\mathcal K_\omega(E)\,dk/dE
          -k(E)\,d\mathcal K_\omega/dE}
        {\mathcal K_\omega^2(E)+k^2(E)},\\
\frac{dk}{dE}&=\frac{\mu}{(\hbar c)^2k}.
\end{aligned}
\label{eq:app_derivative_angle}
\end{equation}
The first line gives radians per MeV; multiplying by \(180/\pi\)
yields approximately \(-24.3839~\mathrm{deg\,MeV}^{-1}\) at the
benchmark energy. Accordingly,
\begin{equation}
\sigma_{\delta,\rm deg}
=\frac{180}{\pi}
 \left|\frac{d\delta_{0,\omega}^{\rm trap}}{dE}\right|_{E=E_c}
 \sigma_E.
\label{eq:app_sigma_angle}
\end{equation}
Here the derivative on the right is in radians per MeV. Because
\(k(E_c)\) and \(\mathcal K_\omega(E_c)\) depend on the same
fitted energy, they are varied together in
Eq.~\eqref{eq:app_derivative_angle}, not propagated as independent
uncertain inputs. No interaction-parameter or extrapolation
uncertainties are included in the tabulated finite-shot errors.

The smooth effective-range reconstruction and threshold relation are
specified in Eqs.~\eqref{eq:app_trap_extrapolation} and
\eqref{eq:app_ere}. They describe the subsequent analysis requiring
additional spectra; no confinement-extrapolation coefficients or
threshold parameters are fitted to the single-level dataset here.

\FloatBarrier

\end{document}